\PassOptionsToPackage{unicode}{hyperref}
\PassOptionsToPackage{hyphens}{url}
\documentclass[
  12pt,
  oneside,
  a4paper,
  fleqn]{article}
\usepackage{amsmath,amssymb}
\usepackage{lmodern}
\usepackage{iftex}
\ifPDFTeX
  \usepackage[T1]{fontenc}
  \usepackage[utf8]{inputenc}
  \usepackage{textcomp} 
\else 
  \usepackage{unicode-math}
  \defaultfontfeatures{Scale=MatchLowercase}
  \defaultfontfeatures[\rmfamily]{Ligatures=TeX,Scale=1}
\fi
\IfFileExists{upquote.sty}{\usepackage{upquote}}{}
\IfFileExists{microtype.sty}{
  \usepackage[]{microtype}
  \UseMicrotypeSet[protrusion]{basicmath} 
}{}
\usepackage{xcolor}

\usepackage[margin=2cm]{geometry}
\usepackage{longtable,booktabs,array}
\usepackage{calc} 
\usepackage{etoolbox}
\makeatletter
\patchcmd\longtable{\par}{\if@noskipsec\mbox{}\fi\par}{}{}
\makeatother
\IfFileExists{footnotehyper.sty}{\usepackage{footnotehyper}}{\usepackage{footnote}}
\makesavenoteenv{longtable}
\usepackage{graphicx}
\makeatletter
\def\maxwidth{\ifdim\Gin@nat@width>\linewidth\linewidth\else\Gin@nat@width\fi}
\def\maxheight{\ifdim\Gin@nat@height>\textheight\textheight\else\Gin@nat@height\fi}
\makeatother
\setkeys{Gin}{width=\maxwidth,height=\maxheight,keepaspectratio}
\makeatletter
\def\fps@figure{htbp}
\makeatother
\usepackage{apacite}
\usepackage{natbib}

\usepackage{float}
\usepackage{amsfonts}
\usepackage{xcolor,graphicx}  
\usepackage{float}
\usepackage{amsthm}
\usepackage{mathrsfs}
\usepackage{wrapfig,lipsum}

\usepackage{tablefootnote}
\PassOptionsToPackage{hyphens}{url}\usepackage{hyperref}

\theoremstyle{definition}

\usepackage[flushleft]{threeparttable}
\allowdisplaybreaks

\newcommand\independent{\protect\mathpalette{\protect\independenT}{\perp}}
    \def\independenT#1#2{\perp\!\!\!\perp}

\usepackage[singlelinecheck=false]{caption}

\usepackage{multirow}
\usepackage{longtable}
\usepackage{booktabs}
\usepackage{amsmath}
\usepackage{bm}
\usepackage{mathrsfs}
\usepackage[italic,noendash]{mathastext}
\usepackage{booktabs}
\usepackage{longtable}
\usepackage{array}
\usepackage{multirow}
\usepackage{wrapfig}
\usepackage{float}
\usepackage{colortbl}
\usepackage{pdflscape}
\usepackage{tabu}
\usepackage{threeparttable}
\usepackage{threeparttablex}
\usepackage[normalem]{ulem}
\usepackage{makecell}
\usepackage{xcolor}
\ifLuaTeX
  \usepackage{selnolig}  
\fi
\usepackage[]{natbib}
\IfFileExists{bookmark.sty}{\usepackage{bookmark}}{\usepackage{hyperref}}
\IfFileExists{xurl.sty}{\usepackage{xurl}}{} 
\hypersetup{
  hidelinks,
  pdfcreator={LaTeX via pandoc}}

\author{}
\date{\vspace{-2.5em}}

\begin{document}

\begin{flushleft}
\vspace{0.8cm}
{ \Large{Sensitivity Analysis of Inverse Probability Weighting Estimators of Causal Effects in Observational Studies with Multivalued Treatments}} \\[18pt]

Md Abdul Basit \footnote{Institute of Statistical Research and Training (ISRT), University of Dhaka, Dhaha 1000, Bangladesh},
Mahbub A.H.M. Latif \textsuperscript{1},
Abdus S Wahed \footnote{Department of Biostatistics, Graduate School of Public Health, University of Pittsburgh,  Pittsburgh, PA 15261, USA}
\end{flushleft}
\begin{flushleft}
{\bf \large{Abstract}}
\end{flushleft}

\noindent One of the fundamental challenges in drawing causal inferences from observational studies is that the assumption of no unmeasured confounding is not testable from observed data. Therefore, assessing sensitivity to this assumption's violation is important to obtain valid causal conclusions in observational studies. Although several sensitivity analysis frameworks are available in the casual inference literature, very few of them are applicable to observational studies with multivalued treatments. To address this issue, we propose a sensitivity analysis framework for performing sensitivity analysis in multivalued treatment settings. Within this framework, a general class of additive causal estimands has been proposed. We demonstrate that the estimation of the causal estimands under the proposed sensitivity model can be performed very efficiently. Simulation results show that the proposed framework performs well in terms of bias of the point estimates and coverage of the confidence intervals when there is sufficient overlap in the covariate distributions. We illustrate the application of our proposed method by conducting an observational study that estimates the causal effect of fish consumption on blood mercury levels.

\hypertarget{introduction}{%
\section{Introduction}\label{introduction}}

Causal inference allows researchers to draw valid causal conclusions based on empirical data.
Most studies in economics, psychology, social sciences, public health, and other disciplines aim at estimating
the causal effect of an intervention or treatment on the outcome of interest. Randomized control
trials (RCTs) are perceived as the gold standard for drawing causal inferences. However, it is
often challenging to implement randomized experiments due to ethical concerns or logistical issues
such as high costs. Therefore, researchers often rely on observational studies to study causal
relationships between the treatment and outcome of interest.

In observational studies, systematic differences might exist among different treatment groups due to the lack of randomization, which necessitates a few stringent assumptions (known as
\emph{identifiability conditions}) about the observed data in order to draw
causal conclusions from observational studies. Based on these
identifiability conditions, several statistical methods have been
developed and thoroughly studied, including g-methods
\citep{horvitz1952generalization, robins1986new, robins1992g},
propensity score \citep{rosenbaum1983central} based methods, and methods
based on machine learning techniques
\citep{robins1994estimation, van2006targeted}.

One of the most common identifiability assumptions in observational
studies is the \emph{strong ignorability} or \emph{no unmeasured confounding}
(NUC) assumption, which implies that the observed covariates include all
common causes or confounders of treatment assignment and outcome of
interest. This assumption is fundamentally unverifiable from
observed data and violation of this assumption may induce substantial bias
in the causal effect estimates. The usefulness of the causal conclusions and subsequent
inferences drawn from observational studies critically rely on the
NUC assumption. Therefore, it is important that a
sensitivity analysis of the estimated causal effects is conducted for correct
interpretation of the results and accurate statement of the causal
conclusions in observational studies.

Sensitivity analysis is widely recognized as a critical step in an observational study. It refers to the approaches that evaluate how the causal estimates and related efficiency measures are affected by the violation of the NUC assumption. Many sponsors and regulatory agencies require that a sensitivity analysis be conducted as part of studies reporting causal conclusions from observational studies. For example, sensitivity analysis is a prerequisite in the Patient-Centered Outcome Research Institute (PCORI) methodology standards for handling missing data (See standard MD-4 in \url{https://bit.ly/3myPhe3}). Sensitivity analysis is one of the five essential steps that need to be conducted before reporting causal effects estimated from an observational study according to the guidelines for statistical reporting of observational studies suggested by \citet{bind2021importance}.

The first formal sensitivity analysis was conducted by \citet{cornfield1959smoking} in order to assess the robustness of the causal relationship between smoking and lung cancer. However, the initial framework of sensitivity analysis did not take a sampling
variation into account and applies only to dichotomous outcomes.
Rosenbaum and colleagues later overcame these shortcomings in a
series of pioneering research articles
\citep{rosenbaum1987sensitivity, gastwirth1998dual, rosenbaum2002covariance}.
\citet{brumback2004sensitivity} proposed an approach in which the
average difference between treatment groups within each stratum formed
by the covariates is used to assess the impact of unmeasured
confounding.
\citet{zhao2019sensitivity} proposed a nonparametric framework of
sensitivity analysis for inverse probability weighting estimators of
causal effects using a marginal sensitivity model initially considered
by \citet{tan2006distributional}. For more sensitivity analysis frameworks in causal inference, readers are referred to \citet{imbens2003sensitivity}, \citet{altonji2005selection}, \citet{arah2008bias}, \citet{shen2011sensitivity}, \citet{carnegie2016assessing}, \citet{dorn2022sharp},
among others. However, all of these sensitivity analysis frameworks apply only to binary treatments.

Observational studies with more than two treatment levels are frequently encountered in practice where the causal effect of a multivalued treatment is of interest. Vector matching \citep{lopez2017estimation}, regression adjustment \citep{linden2016estimating}, and inverse probability weighting \citep{mccaffrey2013tutorial} have been proposed for the estimation of causal effects in multivalued treatment settings. Nevertheless, little has been studied about how sensitivity analysis can be performed in multivalued treatment settings. \citet{bonvini2022sensitivity} proposed a sensitivity analysis framework for marginal structural models (MSMs) encompassing three sensitivity models: propensity-based, outcome-based, and subset confounding. Their framework is applicable to multivalued as well as continuous and time-varying treatments. \citet{zheng2021copula} has also recently proposed a  copula-based sensitivity analysis framework for the multiple treatment setting. In the presence of unmeasured confounding, their framework can be used to characterize bounds for causal effects compatible with the observed data.

In this article, we extend the sensitivity analysis framework of \citet{zhao2019sensitivity} for binary treatments to the multivalued treatment setting using the notion of generalized propensity scores. We also construct a general class of additive causal estimands, which encompasses all causal estimands of interest that can be expressed as linear combinations of the average potential outcomes for each treatment level, e.g., the pairwise average treatment effects. We show that the estimation and sensitivity analysis of causal effects for the proposed class can be performed efficiently using the percentile bootstrap approach proposed by \citet{zhao2019sensitivity}.

The remainder of the paper is organized as follows. Section
\ref{sec:binary_treatments} provides a brief overview of the
observational study problem with binary treatments and the marginal
sensitivity model proposed by \citet{zhao2019sensitivity}. Section
\ref{sec:proposed_framework} describes our proposed sensitivity
analysis framework for observational studies with multivalued
treatments. Section \ref{sec:simulation_study} reports the results from a simulation
study to assess the performance of the proposed method. An illustration of sensitivity analysis in multivalued treatment settings is provided in Section \ref{sec:application} that studies the causal effect of fish consumption on blood mercury levels. Section \ref{sec:discussion} concludes the paper. 


\hypertarget{sensitivity-analysis-in-the-binary-treatment-setting}{%
\section{Sensitivity Analysis in the Binary Treatment Setting}\label{sensitivity-analysis-in-the-binary-treatment-setting}}

\label{sec:binary_treatments}

\hypertarget{the-potential-outcomes-framework}{%
\subsection{The Potential Outcomes Framework}\label{the-potential-outcomes-framework}}

\label{sec:binary_notations}
Suppose, \((A_{1}, \bm{X}_{1}, Y_{1})\), \((A_{2}, \bm{X}_{2}, Y_{2}), \ldots,(A_{n}, \bm{X}_{n}, Y_{n})\) denote data points observed from \(n\) i.i.d. units in an observational study. For each unit \(i \in [n]\), \(A_{i}\) is a binary treatment indicator (1, if treated, and 0, if control), \(\bm{X}_{i} \in \mathscr{X} \subset \mathbb{R}^{d}\) is a vector of measured confounders, and \(Y_i\) is the observed binary outcome. Moreover, let \(Y_{i}(0)\) and \(Y_{i}(1)\) denote the potential outcomes corresponding to treatment levels 0 and 1, respectively, for each unit \(i \in [n]\). Under the stable unit treatment value assumption (SUTVA) \citep{rubin1978bayesian}, the observed outcomes can be defined in terms of potential outcomes as \(Y_{i}=Y_{i}(A_{i})=A_{i} Y_{i}(1) + (1-A_{i}) Y_{i}(0)\). In order to estimate causal effects, we make the following identifying assumptions:

\begin{enumerate}
  \item There is no unmeasured confounding (NUC), i.e., $(Y(0), Y(1)) \independent A \,\vert\, \bm{X}.$ In other words, the set of observed covariates, $\bm{X}_{i}$ includes all common causes of $A_i$ and $Y_i$. (strong ignorability or NUC)
  \item Each unit has a non-zero probability of receiving the treatment. That is, $0 < \mathbb{P}_0(A=1 \,\vert\, X=x) < 1, \thickspace \forall \thickspace A_i$. (positivity or overlap)
\end{enumerate}

In observational studies with a binary treatment, the average treatment effect (ATE), defined as the average difference between the potential outcomes of the treatment and the control group, is one of the most commonly used causal estimands of interest. Estimation of ATE in observational studies heavily relies on the estimation of propensity scores \citep{rosenbaum1983central}, which is defined as the conditional probability of receiving the treatment given the observed confounders.

\hypertarget{the-marginal-sensitivity-model}{%
\subsection{The Marginal Sensitivity Model}\label{the-marginal-sensitivity-model}}

\citet{zhao2019sensitivity} developed a sensitivity analysis framework that can be used for smooth estimators of causal effects, e.g., the inverse probability weighting (IPW) estimators of ATE and ATT in observational studies with a binary treatment. Before defining the marginal sensitivity model, let us assume an unmeasured confounder \(U\) that sums up all unmeasured confounding. \citet{robins2002covariance} suggested that the variable \(U\) can be considered as any of the potential outcomes while defining the conditional probabilities of receiving treatments, i.e., propensity scores. Following Robins's suggestion,
let us denote the complete propensity score (conditional on both observed and unobserved confounders) by \(e_{a}(\bm{x}, y)=\mathbb{P}_{0}\{A=a \,\vert\, \bm{X}=\bm{x}, Y(a)=y\}\) for \(a \in \{0,1\}\) and the observed propensity score (conditional on observed confounders only) by \(e_{0}(\bm{x}) = \mathbb{P}_{0}\{A=1 \,\vert\,\bm{X}=\bm{x}\}\). Then, the marginal sensitivity model assumes that
\begin{align}
    \frac{1}{\Lambda} \leqslant \operatorname{OR}\big\{e_{a}(\bm{x}, y),
    e_{\bm{\beta}_{0}}(\bm{x})\big\} \leqslant \Lambda, \thickspace
    \forall \thickspace \bm{x} \in \mathscr{X}, y \in \mathbb{R},
    a \in\{0,1\}, \label{eq:prIIobs}
\end{align}
where \(\operatorname{OR}(p_1,p_2) = [p_1/(1-p_1)]/[p_2/(1-p_2)]\) is the odds ratio, \(e_{\beta_0}(\bm{x})\) is a parametric considered to model the propensity score \(e_0(\bm{x})\) using observed confounders, and \(\Lambda \geqslant 1\) is a fixed sensitivity parameter that quantifies the magnitude of unmeasured confounding by the odds ratio of the complete propensity score \(e_a(\bm{x},y)\) and the observed propensity score \(e_{\beta_0}(\bm{x})\) for \(a \in \{0,1\}\).
The marginal sensitivity model can also be defined nonparametrically by replacing \(e_{\beta_0}(\bm{x})\) with a nonparametric model \(e_{0}(\bm{x})\) in the sensitivity model defined in \eqref{eq:prIIobs}.

\hypertarget{sensitivity-analysis-in-the-multivalued-treatment-setting}{%
\section{Sensitivity Analysis in the Multivalued Treatment Setting}\label{sensitivity-analysis-in-the-multivalued-treatment-setting}}

\label{sec:proposed_framework}

Consider an observational study with a total of \(J\) treatment levels. We observe i.i.d.
\((A_{i}, \bm{X}_{i}, Y_{i})_{i=1}^{n}\), where
\(A_{i} \in \mathcal{A} = \{1, 2, \ldots , J \}\) with the set of
potential outcomes \(\mathcal{Y}_{i}=\{Y_{i}(1), Y_{i}(2), \ldots, Y_{i}(J)\}\) and
\(\bm{X}_{i} \in \mathscr{X} \subset \mathbb{R}^{d}\) is a vector of
measured confounders for each subject \(i \in [n]\). Moreover, let
\(D_{i}(a)\) be the treatment indicator for \(a \in \mathcal{A}\), such that
\[
  D_{i}(a)=\bigg\{\begin{array}{ll}
  1 & \text { if } A_{i}=a, \\
  0 & \text { otherwise. }
  \end{array}
\]
Then, under SUTVA, the observed outcome \(Y_i\) can be defined in terms of potential outcomes as \(Y_i = \sum_{a=1}^{J} D_i(a) Y_i(a)\) for each subject \(i \in [n]\).

The identifiability assumptions in the multivalued treatment setting are also similar to those in the binary treatment setting. We assume the overlap assumption, which implies that \(\mathbb{P}_{0}(A=a \vert \bm{X}=\bm{x})>0\) for all \(a \in \mathcal{A}\). However, instead of assuming strong ignorability, we assume a weaker version of it, namely, the \emph{weak ignorability} assumption \citep{imbens2000role}:
\[
  D(a) \independent Y(a) \thinspace \vert \thinspace \bm{X}, \thickspace \forall \thinspace a \in \mathcal{A}.
\]
The weak ignorability assumption implies that the treatment \(a \in \mathcal{A}\) assigned to each subject \(i \in [n]\) is independent of the corresponding potential outcome \(Y_i(a)\), instead of the entire set of potential outcomes \(\mathcal{Y}_{i}\). \citet{imbens2000role} showed that an unbiased estimator of the average potential outcomes, \(\mathbb{E}_{0}\big[Y(a)\big]\) for all \(a \in \mathcal{A}\), can be obtained under the weak ignorability assumption. In order to do so, the notion of the propensity score for binary treatments is extended to the multivalued treatment setting by introducing the \emph{generalized propensity score} (GPS). The GPS is defined as the conditional probability of receiving a particular level of the treatment given the measured confounders:
\[
  r_a(\bm{x}) \equiv \mathbb{P}_{0}(A=a \vert \bm{X}=\bm{x}).
\]
For each subject \(i \in [n]\), we obtain a vector of GPSs \(\bm{R(X_i)} = (r_1(\bm{X}_i), \allowbreak \ldots, r_J(\bm{X}_i))\). Using the concept of GPS, an estimand of the average potential outcome at any specific treatment level \(a \in \mathcal{A}\) can be obtained using inverse probability weighting as follows:
\begin{equation}
  m(a) = \mathbb{E}_0[Y(a)] = \mathbb{E}_0\Bigg[\frac{Y \thinspace D(a)}{r_a(\bm{X})}\Bigg].
  \label{eq:ipw_multivalued}
\end{equation}

\hypertarget{the-causal-estimand-of-interest}{%
\subsection{The Causal Estimand of Interest}\label{the-causal-estimand-of-interest}}


In our proposed framework, we consider a general class of additive causal estimands for the multivalued treatment setting. This class of estimands is based on the generalized propensity scores and is defined as linear contrasts of the average potential outcomes \(m(a)\) with the contrast vector \(\bm{c}=(c_1, \ldots, c_J)'\) as
\begin{equation}
  \tau(\bm{c}) = \sum_{a = 1}^{J} c_a m(a) = \sum_{a = 1}^{J} c_a \mathbb{E}_0\Bigg[\frac{Y \thinspace D(a)}{r_a(\bm{X})}\Bigg].
  \label{eq:contrasts}
\end{equation}

The most commonly used estimands of interest in the multivalued treatment setting usually contrast all treatments using $J \choose 2$ simultaneous pairwise comparisons. For any pair of treatments \(\{k,l\} \in \mathcal{A}\), the ATE can be defined using the causal estimand in Equation \eqref{eq:contrasts}. For instance, in an observational study with three treatment levels, the ATE between treatment level \(1\) and \(3\) corresponds to the causal estimand \(\tau(\bm{c})\) with the contrast vector \(\bm{c} = (1,0,-1)'\). Besides pairwise ATEs, any other causal estimand of interest that can be represented as a linear combination of the avarage potential outcomes \(m(a)'s\) can be defined using the causal estimand defined in Equation \eqref{eq:contrasts}.

\hypertarget{the-marginal-sensitivity-model-for-multivalued-treatments}{%
\subsection{The Marginal Sensitivity Model for Multivalued Treatments}\label{the-marginal-sensitivity-model-for-multivalued-treatments}}

Suppose, we use a parametric model \(r_{a,\beta_0}(\bm{x})\) to model the GPS \(r_a(\bm{x})\) for any \(a \in \mathcal{A}\) using observed confounders. Following \citet{zhao2019sensitivity}, for a fixed \(\Lambda \geqslant 1\), the marginal sensitivity model, extended for the multivalued treatment setting, assumes that the complete GPS, \(r_a(\bm{x}, y) = \mathbb{P}_{0}(A=a \vert \bm{X}=\bm{x}, Y(a) = y)\) satisfies
\begin{align}
    \frac{1}{\Lambda} \leqslant \mathrm{OR}\big\{r_a(\bm{x}, y),
    r_{a,\beta_0}(\bm{x}) \big\} \leqslant \Lambda, \thickspace \forall \thickspace 
    \bm{x} \in \mathscr{X},a \in \mathcal{A},y \in \mathbb{R}.
    \label{eq:parasensem}
\end{align}
That is, the marginal sensitivity model for multivalued treatments measures the degree of violation to the NUC assumption by the odds ratio of the true GPS (conditional on both measured and unmeasured confounders) and the observed GPS (conditional on measured confounders only) for all levels $a \in \mathcal{A}$ of the multivalued treatment. As in the binary treatment setting, specification of a marginal sensitivity model for the multivalued treatments depends on a single sensitivity parameter $\Lambda$. Moreover, the sensitivity model \eqref{eq:parasensem} can also be defined nonparametrically using a nonparametric model \(r_{a}(\bm{x})\) in \eqref{eq:parasensem} instead of \(r_{a,\beta_0}( \bm{x})\).

Let us define the logit transforms of \(r_{a,\beta_0}(\bm{x})\) and \(r_a(\bm{x}, y)\) as
\begin{align*}
    g_{a,\beta_0}(\bm{x}) = \operatorname{logit}
    \big(r_{a,\beta_0}(\bm{x})\big) \;\;\; \text { and }
    \;\;\; g_a(\bm{x}, y) = \operatorname{logit}\big(r_a(\bm{x}, y)\big),
\end{align*}
and let \(h_a(\bm{x}, y) = g_{a,\beta_0}(\bm{x}) - g_a(\bm{x}, y)\) for all \(a \in \mathcal{A}\) be the logit-scale difference between the observed and the true GPSs. Then, it can be shown that the sensitivity model \eqref{eq:parasensem} is equivalent to assuming \(h_a \in \mathcal{H}(\lambda)\), where
\begin{align*}
    \mathcal{H}(\lambda)&=\big\{h_a: \mathscr{X} \times \mathbb{R}
    \rightarrow \mathbb{R} \text { and } -\lambda \leqslant h_a \leqslant
    \lambda,\;a\in \mathcal{A}\big\},
\end{align*}
and \(\lambda = \operatorname{log}(\Lambda)\). For a fixed sensitivity parameter \(\lambda \geq 0\), \(h_a(\bm{x}, y)\) will be considered as the sensitivity model from now on.

In a sensitivity analysis, the goal of a researcher is to find out the magnitude of unmeasured confounding, quantified by the sensitivity parameter $\lambda$, that would be needed to invalidate the causal conclusions obtained from a study that assumes no unmeasured confounding. A study is considered sensitive to unmeasured confounding if a small value of $\lambda$ renders the causal effects estimated under the NUC assumption insignificant. Conversely, a study is insensitive to unmeasured confounding if a very large value of $\lambda$ is required to nullify the estimated causal effects. However, determining which values of $\lambda$ can be considered ``large values"  is not straightforward. Calibration refers to approaches that enable researchers to compare the magnitude of unmeasured confounding with the confounding from the observed confounders, leading to a more meaningful interpretation of the sensitivity parameters. Several papers have proposed calibration techniques for different sensitivity analysis frameworks \citep{hsu2013calibrating, zhang2020calibrated, franks2019flexible, cinelli2020making}.

\hypertarget{estimation-of-the-causal-estimands-under-the-marginal-sensitivity-model}{%
\subsection{Estimation of the Causal Estimands under the Marginal Sensitivity Model}\label{estimation-of-the-causal-estimands-under-the-marginal-sensitivity-model}}

In order to incorporate the sensitivity model into the estimation of the causal estimand \eqref{eq:contrasts}, we define the shifted GPSs following \citet{zhao2019sensitivity} as
\begin{align}
    r^{(h_a)}(\bm{x}, y)=\big[1+\exp \big(h_a(\bm{x}, y)-g_{a,\beta_0}(\bm{x})\big)\big]^{-1}.
    \label{eq:shifted_gps}
\end{align}
Using the definition of \(h_a(\bm{x}, y)\) and \(g_{a,\beta_0}(\bm{x})\), it is easy to observe that \(r^{(h_a)}(\bm{x}, y) = r_a(\bm{x}, y)\).

Based on the weak ignorability and overlap assumptions, we define the shifted estimand of the average potential outcomes \(m(a)\) under a specified sensitivity model \(h_a \in \mathcal{H}(\lambda)\) as
\begin{align}
m^{(h_a)}(a) = \mathbb{E}_0\Bigg[\frac{Y \thinspace D(a)}{r_a(\bm{X})}\Bigg]
= \mathbb{E}_0\Bigg[\frac{Y \thinspace D(a)}{r_a(\bm{X}, Y)}\Bigg]
= \mathbb{E}_0\Bigg[\frac{Y \thinspace D(a)}{r^{(h_a)}(\bm{X}, Y)}\Bigg].
\label{shifted_estimator}
\end{align}
Consequently, the shifted causal estimand of interest under a specified sensitivity model \(h_a\) becomes
\begin{align}
  \tau^{(h_a)}(\bm{c}) = \sum_{a = 1}^{J} c_a m^{(h_a)}(a).
  \label{eq:shifted_contrasts}
\end{align}

The set \(\big\{ \tau^{(h_a)}(\bm{c}): h_a(\bm{x}, y) \in \mathcal{H}(\lambda) \big\}\) will be referred to as the partially identified region of \(\tau(\bm{c})\) under a postulated sensitivity model \(h_a \in \mathcal{H}(\lambda)\). That is, instead of a single point estimate, we will obtain an interval of point estimates under a sensitivity model \(h_a\).

In order to obtain an estimate of the shifted estimand \(\tau^{(h_a)}(\bm{c})\), the shifted average potential outcomes \(m^{(h_a)}(a)\) are needed to be estimated for all \(a \in \mathcal{A}\) in Equation \eqref{eq:shifted_contrasts}. This also requires estimates of the parametric GPS models \(r_{a,\beta_0}(\bm{x})\) for all \(a \in \mathcal{A}\) for estimating \(m^{(h_a)}(a)\)'s. In the parametric modeling approach, the common choice for the estimation of GPSs is the multinomial logistic regression model. However, if the treatment levels have a logical ordering, ordinal logistic regression models can also be used as a natural alternative.

Given a sensitivity model \(h_a\), the \emph{stabilized IPW} (SIPW) estimators, obtained by normalizing the inverse probability weights, are defined as:

\begin{align}
    \hat{m}^{(h_a)}&=\Bigg\{\frac{1}{n} \sum_{i=1}^{n} 
    \frac{D_{i}(a)}{\hat{r}^{(h_a)}\big(\bm{X}_{i}, Y_{i}\big)}\Bigg\}^{-1} 
    \Bigg\{\frac{1}{n} \sum_{i=1}^{n} 
    \frac{D_{i}(a) Y_{i}}{\hat{r}^{(h_a)}\big(\bm{X}_{i}, Y_{i}\big)} \Bigg\} \nonumber \\
    &=\frac{\sum_{i=1}^{n} D_{i}(a)Y_i\big[1+\operatorname{exp}\big(h_a(\bm{X}_i,Y_i)
    -g_{a,\hat{\beta}_0}(\bm{X}_i)\big)\big]}{\sum_{i=1}^{n}D_{i}(a)\big
    [1+\operatorname{exp}\big(h_a(\bm{X}_i,Y_i)-g_{a,\hat{\beta}_0}(\bm{X}_i)\big)\big]},
\label{eq:sipw-hat}
\end{align}
where \(r_{a,\hat{\beta}_0}(\bm{x})\) is an estimate of \(r_{a,\beta_0}(\bm{x})\) and \(g_{a,\hat{\beta}_0}(\bm{X}_i) = \operatorname{logit}(r_{a,\hat{\beta}_0}(\bm{x}))\).
\citet{zhao2019sensitivity} showed that estimation problems similar to the one in Equation \eqref{eq:sipw-hat} can be solved very efficiently in partice by converting it to a linear fractional programming problem. Accordingly, Equation \eqref{eq:sipw-hat} is expressed as the following linear fractional programming problem:
\begin{equation}
\begin{aligned}
    \text{minimize or maximize} &\;\;\;\; \frac{\sum_{i=1}^{n} D_{i}(a)Y_i
    \big[1+z_i\operatorname{exp}\big(-g_{a,\hat{\beta}_0}(\bm{X}_i)\big)\big]}
    {\sum_{i=1}^{n}D_{i}(a)\big[1+z_i\operatorname{exp}\big(-g_{a,\hat{\beta}_0}
    (\bm{X}_i)\big)\big]}\\
    \text{subject to} &\;\;\;\; z_{i} \in\big[\Lambda^{-1}, \Lambda\big],
    1 \leqslant i \leqslant n,
\end{aligned}
\label{eq:lfpI}
\end{equation}
where \(z_{i}=\exp \big\{h_a\left(\bm{X}_{i}, Y_{i}\right)\big\}\) for \(i \in[n]\).

The optimization problem in \eqref{eq:lfpI} can further be simplified by converting it to a linear programming problem (LPP) using Charnes-Cooper transformation \citep{charnes1962programming} and can be solved efficiently. Furthermore, under a specified sensitivity model \(h_a\), an asymptotic \(100(1 - \alpha)\%\) confidence interval for the partially identified region of the causal estimand \(\tau(\bm{c})\) can be obtained using a percentile bootstrap approach suggested by \citet{zhao2019sensitivity}. We leave out the details for brevity but would like to note that at most \(J\) optimization problems similar to \eqref{eq:lfpI} need to be solved to estimate the partially identified region of \(\tau(\bm{c})\) with a general contrast vector \(\bm{c}=(c_1, \ldots, c_J)'\). Moreover, these optimization problems are needed to be re-calculated in every bootstrap sample to obtain the corresponding percentile bootstrap interval. However, \citet{zhao2019sensitivity} demonstrated that these computations are extremely efficient, and therefore, involve moderate computational cost.

\hypertarget{simulation-study}{%
\section{Simulation Study}\label{simulation-study}}

\label{sec:simulation_study}

We evaluate the performance of the proposed framework of sensitivity
analysis by simulating an observational study with three treatment
levels. Three covariates are generated as \(X_{i1} \sim \operatorname{Bernoulli}(0.5)\),
\(X_{i2} \sim \operatorname{U}(-1, 1)\), and \(X_{i3} \sim \operatorname{N}(0, 0.5)\). For \(i \in[n]\),
the covariate vector is \(\bm{X}_{i} = (1, X_{i1}, X_{i2}, X_{i3}) ^ {T}\).
The treatment assignment mechanism follows a multinomial distribution
\[
\big(D_{i}(1), D_{i}(2), D_{i}(3)\big) \;\big\vert\; \bm{X}_{i} \sim 
\operatorname{Multinom}\big(r_{1}(\bm{X}_{i}), r_{2}(\bm{X}_{i}), r_{3}(\bm{X}_{i})\big),
\]
where \(D_{i}(a)\) is the treatment indicator and
\[
r_{a}(\bm{X}_{i}, Y_i) = r_{a}(\bm{X}_{i}) = \frac{\exp \big(\bm{X}_{i}^{T} \beta_{a}\big)}{\sum_{a^\prime=1}^{3} \exp \big(\bm{X}_{i}^{T} \beta_{a^\prime}\big)} 
\]
is the true GPS for \(a \in \{1,2,3\}\) with \(\beta_1 = (0, 0, 0, 0)^T\), \(\beta_2 = k_2 \times (0, 1, 1, 1)^T\),
and \(\beta_3 = k_3 \times (0, 1, 1, -1)^T\). In order to assess the influence of the degree of overlap on the point estimates and confidence intervals of our proposed framework, we consider
two simulation scenarios. In the first scenario (Scenario-I),
we set \((k_2, k_3) = (0.1, -0.1)\) to simulate a scenario with adequate overlap in
the covariates, and in the second scenario (Scenario-II), we set \((k_2, k_3) = (3, 3)\) to induce lack of overlap.
The potential outcomes are generated from the following multinomial distribution

\[
\big(Y_{i}(1), Y_{i}(2), Y_{i}(3)\big) \;\big\vert\; \bm{X}_{i} \sim 
\operatorname{Multinom}\big(p_{Y_1}(\bm{X}_{i}), p_{Y_2}(\bm{X}_{i}), p_{Y_3}(\bm{X}_{i})\big),
\]
where

\[
p_{Y_a}(\bm{X}_{i}) = \mathbb{P}(Y(a) = 1 \big \vert \bm{X}_{i}) = \frac{\exp \big(\bm{X}_{i}^{T} \delta_{a}\big)}{\sum_{a^\prime=1}^{3} \exp \big( \bm{X}_{i}^{T} \delta_{a^\prime}\big)} 
\]
with \(\delta_1 = (1, 1, 1, 1)^T\),
\(\delta_2 = (1, 1, -1, 1)^T\), and \(\delta_3 = (1, 1, 1, -1)^T\). The observed outcome is then obtained as \(Y_i = \sum_{a=1}^{J} D_i(a) Y_i(a)\) for each subject \(i \in [n]\). For each scenario, we simulate \(1000\) datasets with sample size \(n=750\) and estimate the interval of point
estimates and confidence intervals for the pairwise ATEs using the proposed framework for multivalued treatments. The three pairwise ATEs of interest are denoted by \(\tau_{1,2}\), \(\tau_{1,3}\), and \(\tau_{2,3}\), respectively. We consider six values of the sensitivity
parameter \(\lambda\), namely, \(\lambda = \{0, 0.1, 0.2, 0.5, 1, 2\}\).
The true partially identified intervals are obtained under each simulation scenario using large scale numerical approximations.
Moreover, since the treatment under consideration has three levels, the observed generalized
propensity scores (GPSs) are modeled using multinomial logit regression models.

\begin{table}[!h]

\caption{\label{tab:multivalued-OR}Simulation results for the sensitivity analysis of pairwise ATEs in an observational study with a three treatment levels.\textsuperscript{\dag}}
\centering
\resizebox{\linewidth}{!}{
\begin{threeparttable}
\begin{tabular}[t]{ccrrrrrrrr}
\toprule
\multicolumn{4}{c}{ } & \multicolumn{2}{c}{\% Bias} & \multicolumn{4}{c}{ } \\
\cmidrule(l{3pt}r{3pt}){5-6}
Scenario & ATE & $\lambda$ & $\Lambda$ & Lower & Upper & \makecell[c]{Non-\\ coverage} & \makecell[c]{Partially identified \\ interval} & \makecell[c]{Point estimate \\ interval} & \makecell[c]{Confidence \\ interval}\\
\midrule
 &  & 0.0 & 1.00 & $2.79$ & $2.79$ & 0.103 & $(-0.050, -0.050)$ & $(-0.048, -0.048)$ & $(-0.116, 0.017)$\\

 &  & 0.1 & 1.11 & $3.43$ & $3.01$ & 0.104 & $(-0.108, 0.008)$ & $(-0.107, 0.010)$ & $(-0.171, 0.076)$\\

 &  & 0.2 & 1.22 & $3.40$ & $0.65$ & 0.101 & $(-0.165, 0.067)$ & $(-0.164, 0.067)$ & $(-0.228, 0.133)$\\

 &  & 0.5 & 1.65 & $2.97$ & $0.59$ & 0.115 & $(-0.328, 0.237)$ & $(-0.328, 0.238)$ & $(-0.387, 0.302)$\\

 &  & 1.0 & 2.72 & $6.74$ & $-3.17$ & 0.124 & $(-0.559, 0.488)$ & $(-0.558, 0.489)$ & $(-0.606, 0.543)$\\

 & \multirow{-6}{*}{\centering\arraybackslash $\tau_{1, 2}$} & 2.0 & 7.39 & $6.67$ & $-5.32$ & 0.110 & $(-0.832, 0.801)$ & $(-0.832, 0.802)$ & $(-0.856, 0.830)$\\
\cmidrule{2-10}
 &  & 0.0 & 1.00 & $1.49$ & $1.49$ & 0.105 & $(-0.034, -0.034)$ & $(-0.035, -0.035)$ & $(-0.101, 0.033)$\\

 &  & 0.1 & 1.11 & $2.15$ & $1.12$ & 0.110 & $(-0.093, 0.025)$ & $(-0.093, 0.024)$ & $(-0.161, 0.092)$\\

 &  & 0.2 & 1.22 & $4.10$ & $2.50$ & 0.111 & $(-0.152, 0.083)$ & $(-0.152, 0.084)$ & $(-0.217, 0.150)$\\

 &  & 0.5 & 1.65 & $5.27$ & $0.69$ & 0.111 & $(-0.320, 0.254)$ & $(-0.319, 0.255)$ & $(-0.380, 0.318)$\\

 &  & 1.0 & 2.72 & $7.27$ & $-2.13$ & 0.126 & $(-0.558, 0.502)$ & $(-0.556, 0.502)$ & $(-0.607, 0.556)$\\

 & \multirow{-6}{*}{\centering\arraybackslash $\tau_{1, 3}$} & 2.0 & 7.39 & $9.38$ & $-5.69$ & 0.110 & $(-0.837, 0.808)$ & $(-0.836, 0.808)$ & $(-0.861, 0.837)$\\
\cmidrule{2-10}
 &  & 0.0 & 1.00 & $1.22$ & $1.22$ & 0.092 & $(0.015, 0.015)$ & $(0.016, 0.016)$ & $(-0.052, 0.085)$\\

 &  & 0.1 & 1.11 & $0.26$ & $-0.80$ & 0.101 & $(-0.045, 0.076)$ & $(-0.044, 0.077)$ & $(-0.112, 0.144)$\\

 &  & 0.2 & 1.22 & $2.31$ & $0.46$ & 0.101 & $(-0.106, 0.135)$ & $(-0.105, 0.136)$ & $(-0.173, 0.202)$\\

 &  & 0.5 & 1.65 & $3.40$ & $-0.91$ & 0.093 & $(-0.281, 0.306)$ & $(-0.280, 0.306)$ & $(-0.345, 0.367)$\\

 &  & 1.0 & 2.72 & $4.48$ & $-2.57$ & 0.102 & $(-0.531, 0.546)$ & $(-0.531, 0.547)$ & $(-0.583, 0.596)$\\

\multirow{-18}{*}{\centering\arraybackslash I} & \multirow{-6}{*}{\centering\arraybackslash $\tau_{2, 3}$} & 2.0 & 7.39 & $9.62$ & $-7.57$ & 0.102 & $(-0.827, 0.829)$ & $(-0.827, 0.829)$ & $(-0.851, 0.853)$\\
\cmidrule{1-10}
 &  & 0.0 & 1.00 & $-15.30$ & $-15.30$ & 0.189 & $(-0.044, -0.044)$ & $(-0.084, -0.084)$ & $(-0.261, 0.101)$\\

 &  & 0.1 & 1.11 & $-11.34$ & $-19.21$ & 0.205 & $(-0.106, 0.019)$ & $(-0.140, -0.026)$ & $(-0.307, 0.159)$\\

 &  & 0.2 & 1.22 & $-7.02$ & $-23.34$ & 0.225 & $(-0.167, 0.083)$ & $(-0.194, 0.033)$ & $(-0.350, 0.219)$\\

 &  & 0.5 & 1.65 & $3.27$ & $-35.38$ & 0.274 & $(-0.334, 0.273)$ & $(-0.341, 0.207)$ & $(-0.474, 0.389)$\\

 &  & 1.0 & 2.72 & $16.90$ & $-51.30$ & 0.376 & $(-0.559, 0.548)$ & $(-0.551, 0.475)$ & $(-0.647, 0.620)$\\

 & \multirow{-6}{*}{\centering\arraybackslash $\tau_{1, 2}$} & 2.0 & 7.39 & $30.93$ & $-65.77$ & 0.500 & $(-0.824, 0.852)$ & $(-0.812, 0.813)$ & $(-0.860, 0.869)$\\
\cmidrule{2-10}
 &  & 0.0 & 1.00 & $-15.57$ & $-15.57$ & 0.203 & $(-0.030, -0.030)$ & $(-0.072, -0.072)$ & $(-0.243, 0.101)$\\

 &  & 0.1 & 1.11 & $-12.37$ & $-19.35$ & 0.217 & $(-0.089, 0.031)$ & $(-0.126, -0.017)$ & $(-0.285, 0.162)$\\

 &  & 0.2 & 1.22 & $-8.88$ & $-22.72$ & 0.230 & $(-0.147, 0.092)$ & $(-0.179, 0.037)$ & $(-0.327, 0.218)$\\

 &  & 0.5 & 1.65 & $-0.37$ & $-33.93$ & 0.274 & $(-0.307, 0.276)$ & $(-0.324, 0.211)$ & $(-0.448, 0.386)$\\

 &  & 1.0 & 2.72 & $9.74$ & $-48.42$ & 0.355 & $(-0.525, 0.544)$ & $(-0.525, 0.475)$ & $(-0.617, 0.617)$\\

 & \multirow{-6}{*}{\centering\arraybackslash $\tau_{1, 3}$} & 2.0 & 7.39 & $21.21$ & $-61.01$ & 0.472 & $(-0.798, 0.846)$ & $(-0.790, 0.809)$ & $(-0.840, 0.865)$\\
\cmidrule{2-10}
 &  & 0.0 & 1.00 & $-0.26$ & $-0.26$ & 0.129 & $(0.014, 0.014)$ & $(0.014, 0.014)$ & $(-0.110, 0.135)$\\

 &  & 0.1 & 1.11 & $2.08$ & $-2.59$ & 0.141 & $(-0.037, 0.065)$ & $(-0.035, 0.063)$ & $(-0.157, 0.183)$\\

 &  & 0.2 & 1.22 & $3.60$ & $-5.17$ & 0.154 & $(-0.087, 0.116)$ & $(-0.084, 0.111)$ & $(-0.203, 0.229)$\\

 &  & 0.5 & 1.65 & $8.88$ & $-11.67$ & 0.175 & $(-0.234, 0.264)$ & $(-0.227, 0.254)$ & $(-0.334, 0.363)$\\

 &  & 1.0 & 2.72 & $17.29$ & $-21.99$ & 0.253 & $(-0.455, 0.485)$ & $(-0.441, 0.468)$ & $(-0.532, 0.557)$\\

\multirow{-18}{*}{\centering\arraybackslash II} & \multirow{-6}{*}{\centering\arraybackslash $\tau_{2, 3}$} & 2.0 & 7.39 & $27.26$ & $-34.41$ & 0.335 & $(-0.760, 0.781)$ & $(-0.748, 0.767)$ & $(-0.799, 0.814)$\\
\bottomrule
\end{tabular}
\begin{tablenotes}
\item[\dag] The first four columns represents the simulation settings: Scenario and ATE are the simulation scenarios and pairwise ATEs of interest, respectively; $\lambda$ is the sensitivity parameter and $\Lambda = exp(\lambda)$. The next five columns are the results: respectively percentage average bias (in SD units) in the lower and upper bound of the point estimate interval; the non-coverage rate of the confidence intervals under the proposed framework (the desired non-coverage rate is $10\%$); the partially identified interval of the mean response; the median interval for the point estimates; the median confidence interval.
\end{tablenotes}
\end{threeparttable}}
\end{table}

We simulate \(1000\) datasets for each scenario and estimate the interval of point estimates and construct \(90\%\) confidence intervals for the pairwise ATEs under the proposed sensitivity analysis framework for multivalued treatment settings.
We report the percentage average bias (in SD units) in the lower and upper bounds of the point estimate interval, the non-coverage rate of the confidence interval, the median interval of point estimates, and the median confidence intervals calculated from the \(1000\) simulated datasets in Table \ref{tab:multivalued-OR}.

When there is an adequate overlap (Scenario-I), we observe that the confidence intervals constructed using percentile bootstrap in our proposed framework satisfy the nominal \(10\%\) non-coverage rate for all pairwise ATEs in most cases. However, the confidence intervals suffer from a high non-coverage rate when there is a lack of overlap (Scenario-II), and the non-coverage deteriorates as the sensitivity parameter \(\lambda\) increases. These observations are similar to that of \citet{zhao2019sensitivity} on the performance of their proposed percentile bootstrap confidence intervals considered for the missing data problem. Regarding bias, the SIPW point estimators also perform well when there is adequate overlap (Scenario-I). The average bias of each SIPW estimate remains within \(10\%\) of its standard deviation in almost all cases. However, the magnitude of bias in the point estimates increases due to a lack of overlap (Scenario-II).

We want to point out a final observation that \citet{zhao2019sensitivity} did not report significant bias for SIPW estimates under their sensitivity analysis framework for situations with non-overlapping covariate distributions. In contrast, we measured the average bias in the point estimates as the percentage of corresponding standard deviations and observed that not only the non-coverage rate of the confidence intervals but also the bias of the point estimates increase when there is a lack of covariate overlap. These findings are also supported by several studies pointing out that lack of covariate overlap results in instability in the variance leading to a high non-coverage rate for confidence intervals and high bias in the point estimates in observational studies \citep{li2019addressing, hu2020estimation}.

\hypertarget{application}{%
\section{Application}\label{application}}

\label{sec:application}

In this section, the dataset analyzed in \citet{zhao2019sensitivity} is extended to a multivalued treatment setting to demonstrate the application of the proposed sensitivity analysis framework introduced in Section \ref{sec:proposed_framework}. This dataset is derived from the National Health and Nutrition Examination Survey (NHANES) 2013-14 containing \(1107\) responses from individuals who were at least 18 years old, answered the questionnaire about seafood consumption, and had blood mercury levels measured. It is publicly available in the \texttt{R} package \texttt{CrossScreening} on the Comprehensive \texttt{R} Archive Network. \footnote{Package \texttt{CrossScreening} was removed from the CRAN repository. The dataset can be obtained from the archived version of the package available at \url{https://bit.ly/41jzupx}.} \citet{zhao2019sensitivity} converted the measured fish consumption into a binary treatment variable to illustrate their marginal sensitivity model for binary treatment settings. They defined ``high fish'' consumption as more than 12 servings of fish or shellfish in the previous month and ``low fish'' consumption as no or less than 12 servings of fish. Unlike \citet{zhao2019sensitivity}, we categorized fish consumption into three categories leading to a multivalued treatment with three treatment levels. We defined ``high fish'' consumption as more than 12 servings of fish or shellfish in the previous month, ``low fish'' consumption as 1 to 12 servings of fish, and ``no fish'' consumption as 0 servings of fish in the previous month. This categorization resulted in 234 individuals with ``high fish'' consumption, 328 individuals with ``low fish'' consumption, and 545 individuals with ``no fish'' consumption. Similar to \citet{zhao2019sensitivity}, we considered eight covariates, namely, gender, age, income, whether income is missing, education, race, ever smoked, and the number of cigarettes smoked last month. The outcome variable is \(\log _2\)(total blood mercury) in micrograms per litre.

\begin{table}[!h]
\caption{\label{tab:nhanes}SIPW point estimate intervals and 90\% percentile bootstrap confidence intervals for the pairwise ATEs for different values of the sensitivity parameter $\lambda$. The pariwise ATEs between no vs. low consumption, no vs. high consumption, and low vs. high consumption are denoted by  $\tau_{1, 2}$, $\tau_{1, 3}$, and $\tau_{2, 3}$, respectively.}
\centering
\begin{tabular}[t]{crrrr}
\toprule
Estimand & $\Lambda$ & $\lambda$ & \makecell[c]{Point estimate \\ interval} & \makecell[c]{90\% confidence \\ interval}\\
\midrule
 & 1 & 0.00 & $(0.45, 0.45)$ & $(0.31, 0.58)$\\

 & 1.65 & 0.50 & $(-0.09, 0.99)$ & $(-0.22, 1.14)$\\

 & 2.12 & 0.75 & $(-0.36, 1.25)$ & $(-0.50, 1.40)$\\

 & 2.72 & 1.00 & $(-0.62, 1.51)$ & $(-0.76, 1.68)$\\

 & 4.48 & 1.50 & $(-1.13, 2.01)$ & $(-1.30, 2.23)$\\

\multirow{-6}{*}{\centering\arraybackslash $\tau_{1, 2}$} & 7.39 & 2.00 & $(-1.62, 2.50)$ & $(-1.78, 2.78)$\\
\cmidrule{1-5}
 & 1 & 0.00 & $(2.08, 2.08)$ & $(1.89, 2.27)$\\

 & 1.65 & 0.50 & $(1.44, 2.69)$ & $(1.25, 2.86)$\\

 & 2.12 & 0.75 & $(1.14, 2.98)$ & $(0.93, 3.13)$\\

 & 2.72 & 1.00 & $(0.84, 3.25)$ & $(0.63, 3.41)$\\

 & 4.48 & 1.50 & $(0.29, 3.77)$ & $(0.08, 3.94)$\\

\multirow{-6}{*}{\centering\arraybackslash $\tau_{1, 3}$} & 7.39 & 2.00 & $(-0.24, 4.27)$ & $(-0.46, 4.45)$\\
\cmidrule{1-5}
 & 1 & 0.00 & $(1.63, 1.63)$ & $(1.44, 1.84)$\\

 & 1.65 & 0.50 & $(0.90, 2.34)$ & $(0.69, 2.53)$\\

 & 2.12 & 0.75 & $(0.55, 2.68)$ & $(0.33, 2.85)$\\

 & 2.72 & 1.00 & $(0.21, 3.00)$ & $(-0.02, 3.18)$\\

 & 4.48 & 1.50 & $(-0.42, 3.59)$ & $(-0.67, 3.77)$\\

\multirow{-6}{*}{\centering\arraybackslash $\tau_{2, 3}$} & 7.39 & 2.00 & $(-1.02, 4.17)$ & $(-1.33, 4.39)$\\
\bottomrule
\end{tabular}
\end{table}

We used our proposed framework for multivalued treatments to perform sensitivity analysis for the three pairwise ATEs \(\tau_{1,2}\) (``no fish'' vs.~``low fish'' consumption), \(\tau_{1,3}\) (``no fish'' vs.~``high fish'' consumption), and \(\tau_{2,3}\) (``low fish'' vs.~``high fish'' consumption). Since there is a natural ordering among the fish consumption levels, we fit an ordinal logistic regression model, namely the continuation ratio regression model, to estimate the generalized propensity scores (GPSs) using the chosen observed confounders. The estimated GPSs range from \(0.008\) to \(0.825\) with a mean of \(0.25\) across all treatment levels. Following the findings of our simulation study in Section \ref{sec:simulation_study}, the obtained results should be interpreted carefully as some GPSs are close to 0, which is indicative of a potential lack of overlap in covariate distributions. We obtain 90\% confidence intervals for the pairwise ATEs using \(1000\) bootstrap samples for six values of the sensitivity parameter \(\lambda = \operatorname{log}(\Lambda) = \{0, 0.5, 0.75, 1, 1.5, 2\}\).

\begin{figure}
\centering
\includegraphics{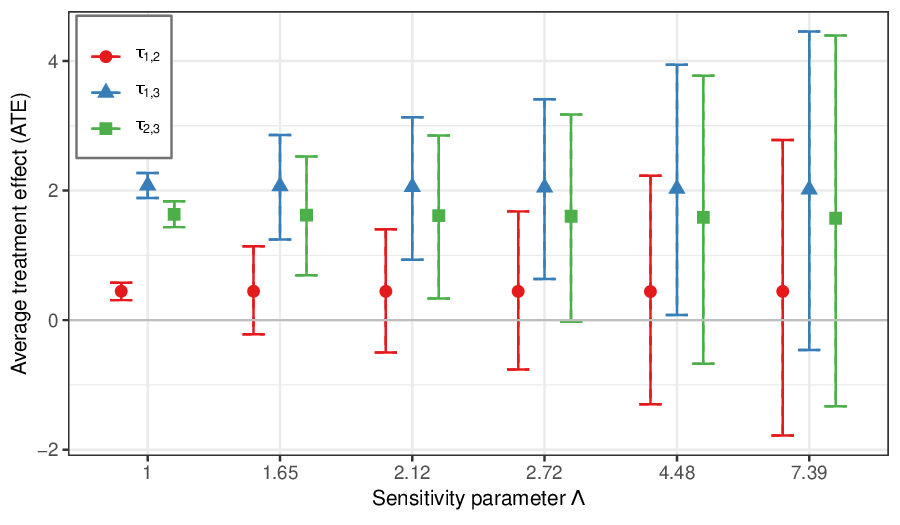}
\caption{\label{fig:unnamed-chunk-2}\label{fig:sens_plot}Graphical representation of the sensitivity analysis results for the three pairwise ATEs. The solid error bars represent the range of point estimates and the dashed error bars (together with the solid error bars) represent the confidence intervals. The circles/triangles/squares are the mid-points of the solid error bars.}
\end{figure}

The results of the sensitivity analysis are presented in Table \ref{tab:nhanes}.
Moreover, Figure \ref{fig:sens_plot} depicts the range of point estimates and confidence intervals for six different sensitivity parameter values. We observe that the pairwise ATE between ``no fish'' and ``high fish'' consumption (\(\tau_{1,3}\)) is significantly different from the null value \(0\) for \(\Lambda <= 4.48\). In contrast, the pairwise ATE between ``low fish'' and ``high fish'' consumption (\(\tau_{2,3}\)) is significantly different from \(0\) for \(\Lambda <= 2.12\). However, the confidence interval for the pairwise ATE between ``no fish'' and ``low fish'' consumption (\(\tau_{1,2}\)) contains the null value 0 for relatively small sensitivity parameter values such as \(\Lambda = 1.65\). These results imply that for the pairwise ATEs between ``no fish'' and ``high fish'' consumption as well as between ``low fish'' and ``high fish'' consumption to be insignificant, the estimated GPSs (given measured confounders only) must deviate substantially from the complete GPSs (given both measured and unmeasured confounders). This can be interpreted as a fairly strong evidence favoring a significant causal effect of fish consumption on individuals' blood mercury levels that are insensitive to unmeasured confounding.

\hypertarget{discussion}{%
\section{Discussion}\label{discussion}}

\label{sec:discussion}

Sensitivity analysis is essential for examining the influence of unmeasured confounding on the causal conclusions derived from observational studies. In this paper, we extended the sensitivity analysis framework of \citet{zhao2019sensitivity} and proposed a sensitivity analysis framework that can be used to appraise the sensitivity of causal estimates to unmeasured confounding in observational studies with multivalued treatments. We employed a general class of additive causal estimands for multivalued treatments and showed that the estimation of the causal estimands and sensitivity analysis of the estimated causal effects could be performed efficiently using statistical methods similar to \citet{zhao2019sensitivity}.

The simulation results presented in Section \ref{sec:simulation_study} suggest that our proposed sensitivity analysis framework performs well in terms of bias in the point estimate intervals and coverage of the confidence intervals when there is an adequate overlap in the covariate distribution among the treatment groups. However, The performance deteriorates when there exists a lack of adequate overlap among the treatment groups. Moreover, an application of the proposed framework in multivalued treatment settings is illustrated by conducting an observational study that estimates the causal effect of fish consumption on blood mercury levels.

There are several potential future directions of our present work. It is possible to make the partially identified intervals and the associated confidence intervals of the proposed framework narrower using the quantile balancing method suggested by \citet{dorn2022sharp}. Besides the IPW estimators of causal effects, the proposed framework could also be extended to incorporate other smooth estimators of causal effects, such as the doubly robust augmented IPW (AIPW) estimators \citep{robins1994estimation} and generalized overlap weighting (GOW) estimators \citep{li2019propensity}. Lastly, the sensitivity parameter of the proposed framework should be calibrated to the measured confounders to interpret the sensitivity analysis results more meaningfully.



\renewcommand\refname{REFERENCES}
  \bibliography{statmed-submission}

\begin{thebibliography}{}

\bibitem[Altonji et~al., 2005]{altonji2005selection}
Altonji, J.~G., Elder, T.~E., and Taber, C.~R. (2005).
\newblock Selection on observed and unobserved variables: Assessing the
  effectiveness of catholic schools.
\newblock {\em Journal of Political Economy}, 113(1):151--184.

\bibitem[Arah et~al., 2008]{arah2008bias}
Arah, O.~A., Chiba, Y., and Greenland, S. (2008).
\newblock Bias formulas for external adjustment and sensitivity analysis of
  unmeasured confounders.
\newblock {\em Annals of Epidemiology}, 18(8):637--646.

\bibitem[Bind and Rubin, 2021]{bind2021importance}
Bind, M.-A.~C. and Rubin, D. (2021).
\newblock The importance of having a conceptual stage when reporting
  non-randomized studies.
\newblock {\em Biostatistics \& Epidemiology}, 5(1):9--18.

\bibitem[Bonvini et~al., 2022]{bonvini2022sensitivity}
Bonvini, M., Kennedy, E., Ventura, V., and Wasserman, L. (2022).
\newblock Sensitivity analysis for marginal structural models.
\newblock {\em arXiv preprint arXiv:2210.04681}.

\bibitem[Brumback et~al., 2004]{brumback2004sensitivity}
Brumback, B.~A., Hern{\'a}n, M.~A., Haneuse, S.~J., and Robins, J.~M. (2004).
\newblock Sensitivity analyses for unmeasured confounding assuming a marginal
  structural model for repeated measures.
\newblock {\em Statistics in medicine}, 23(5):749--767.

\bibitem[Carnegie et~al., 2016]{carnegie2016assessing}
Carnegie, N.~B., Harada, M., and Hill, J.~L. (2016).
\newblock Assessing sensitivity to unmeasured confounding using a simulated
  potential confounder.
\newblock {\em Journal of Research on Educational Effectiveness},
  9(3):395--420.

\bibitem[Charnes and Cooper, 1962]{charnes1962programming}
Charnes, A. and Cooper, W.~W. (1962).
\newblock Programming with linear fractional functionals.
\newblock {\em Naval Research logistics quarterly}, 9(3-4):181--186.

\bibitem[Cinelli and Hazlett, 2020]{cinelli2020making}
Cinelli, C. and Hazlett, C. (2020).
\newblock Making sense of sensitivity: Extending omitted variable bias.
\newblock {\em Journal of the Royal Statistical Society Series B: Statistical
  Methodology}, 82(1):39--67.

\bibitem[Cornfield et~al., 1959]{cornfield1959smoking}
Cornfield, J., Haenszel, W., Hammond, E.~C., Lilienfeld, A.~M., Shimkin, M.~B.,
  and Wynder, E.~L. (1959).
\newblock Smoking and lung cancer: recent evidence and a discussion of some
  questions.
\newblock {\em Journal of the National Cancer Institute}, 22(1):173--203.

\bibitem[Dorn and Guo, 2022]{dorn2022sharp}
Dorn, J. and Guo, K. (2022).
\newblock Sharp sensitivity analysis for inverse propensity weighting via
  quantile balancing.
\newblock {\em Journal of the American Statistical Association}, pages 1--13.

\bibitem[Franks et~al., 2019]{franks2019flexible}
Franks, A., D’Amour, A., and Feller, A. (2019).
\newblock Flexible sensitivity analysis for observational studies without
  observable implications.
\newblock {\em Journal of the American Statistical Association}.

\bibitem[Gastwirth et~al., 1998]{gastwirth1998dual}
Gastwirth, J.~L., Krieger, A.~M., and Rosenbaum, P.~R. (1998).
\newblock Dual and simultaneous sensitivity analysis for matched pairs.
\newblock {\em Biometrika}, 85(4):907--920.

\bibitem[Horvitz and Thompson, 1952]{horvitz1952generalization}
Horvitz, D.~G. and Thompson, D.~J. (1952).
\newblock A generalization of sampling without replacement from a finite
  universe.
\newblock {\em Journal of the American Statistical Association},
  47(260):663--685.

\bibitem[Hsu and Small, 2013]{hsu2013calibrating}
Hsu, J.~Y. and Small, D.~S. (2013).
\newblock Calibrating sensitivity analyses to observed covariates in
  observational studies.
\newblock {\em Biometrics}, 69(4):803--811.

\bibitem[Hu et~al., 2020]{hu2020estimation}
Hu, L., Gu, C., Lopez, M., Ji, J., and Wisnivesky, J. (2020).
\newblock Estimation of causal effects of multiple treatments in observational
  studies with a binary outcome.
\newblock {\em Statistical methods in medical research}, 29(11):3218--3234.

\bibitem[Imbens, 2000]{imbens2000role}
Imbens, G.~W. (2000).
\newblock The role of the propensity score in estimating dose-response
  functions.
\newblock {\em Biometrika}, 87(3):706--710.

\bibitem[Imbens, 2003]{imbens2003sensitivity}
Imbens, G.~W. (2003).
\newblock Sensitivity to exogeneity assumptions in program evaluation.
\newblock {\em American Economic Review}, 93(2):126--132.

\bibitem[Li and Li, 2019]{li2019propensity}
Li, F. and Li, F. (2019).
\newblock Propensity score weighting for causal inference with multiple
  treatments.
\newblock {\em The Annals of Applied Statistics}, 13(4):2389--2415.

\bibitem[Li et~al., 2019]{li2019addressing}
Li, F., Thomas, L.~E., and Li, F. (2019).
\newblock Addressing extreme propensity scores via the overlap weights.
\newblock {\em American journal of epidemiology}, 188(1):250--257.

\bibitem[Linden et~al., 2016]{linden2016estimating}
Linden, A., Uysal, S.~D., Ryan, A., and Adams, J.~L. (2016).
\newblock Estimating causal effects for multivalued treatments: a comparison of
  approaches.
\newblock {\em Statistics in Medicine}, 35(4):534--552.

\bibitem[Lopez et~al., 2017]{lopez2017estimation}
Lopez, M.~J., Gutman, R., et~al. (2017).
\newblock Estimation of causal effects with multiple treatments: a review and
  new ideas.
\newblock {\em Statistical Science}, 32(3):432--454.

\bibitem[McCaffrey et~al., 2013]{mccaffrey2013tutorial}
McCaffrey, D.~F., Griffin, B.~A., Almirall, D., Slaughter, M.~E., Ramchand, R.,
  and Burgette, L.~F. (2013).
\newblock A tutorial on propensity score estimation for multiple treatments
  using generalized boosted models.
\newblock {\em Statistics in medicine}, 32(19):3388--3414.

\bibitem[Robins, 1986]{robins1986new}
Robins, J. (1986).
\newblock A new approach to causal inference in mortality studies with a
  sustained exposure period-application to control of the healthy worker
  survivor effect.
\newblock {\em Mathematical Modelling}, 7(9-12):1393--1512.

\bibitem[Robins, 2002]{robins2002covariance}
Robins, J.~M. (2002).
\newblock [covariance adjustment in randomized experiments and observational
  studies]: Comment.
\newblock {\em Statistical Science}, 17(3):309--321.

\bibitem[Robins et~al., 1992]{robins1992g}
Robins, J.~M., Blevins, D., Ritter, G., and Wulfsohn, M. (1992).
\newblock G-estimation of the effect of prophylaxis therapy for pneumocystis
  carinii pneumonia on the survival of aids patients.
\newblock {\em Epidemiology}, 3(4):319--336.

\bibitem[Robins et~al., 1994]{robins1994estimation}
Robins, J.~M., Rotnitzky, A., and Zhao, L.~P. (1994).
\newblock Estimation of regression coefficients when some regressors are not
  always observed.
\newblock {\em Journal of the American Statistical Association},
  89(427):846--866.

\bibitem[Rosenbaum, 1987]{rosenbaum1987sensitivity}
Rosenbaum, P.~R. (1987).
\newblock Sensitivity analysis for certain permutation inferences in matched
  observational studies.
\newblock {\em Biometrika}, 74(1):13--26.

\bibitem[Rosenbaum et~al., 2002]{rosenbaum2002covariance}
Rosenbaum, P.~R. et~al. (2002).
\newblock Covariance adjustment in randomized experiments and observational
  studies.
\newblock {\em Statistical Science}, 17(3):286--327.

\bibitem[Rosenbaum and Rubin, 1983]{rosenbaum1983central}
Rosenbaum, P.~R. and Rubin, D.~B. (1983).
\newblock The central role of the propensity score in observational studies for
  causal effects.
\newblock {\em Biometrika}, 70(1):41--55.

\bibitem[Rubin, 1978]{rubin1978bayesian}
Rubin, D.~B. (1978).
\newblock Bayesian inference for causal effects: The role of randomization.
\newblock {\em The Annals of Statistics}, 6(1):34--58.

\bibitem[Shen et~al., 2011]{shen2011sensitivity}
Shen, C., Li, X., Li, L., and Were, M.~C. (2011).
\newblock Sensitivity analysis for causal inference using inverse probability
  weighting.
\newblock {\em Biometrical Journal}, 53(5):822--837.

\bibitem[Tan, 2006]{tan2006distributional}
Tan, Z. (2006).
\newblock A distributional approach for causal inference using propensity
  scores.
\newblock {\em Journal of the American Statistical Association},
  101(476):1619--1637.

\bibitem[Van Der~Laan and Rubin, 2006]{van2006targeted}
Van Der~Laan, M.~J. and Rubin, D. (2006).
\newblock Targeted maximum likelihood learning.
\newblock {\em The International Journal of Biostatistics}, 2(1).

\bibitem[Zhang and Small, 2020]{zhang2020calibrated}
Zhang, B. and Small, D.~S. (2020).
\newblock A calibrated sensitivity analysis for matched observational studies
  with application to the effect of second-hand smoke exposure on blood lead
  levels in children.
\newblock {\em Journal of the Royal Statistical Society Series C: Applied
  Statistics}, 69(5):1285--1305.

\bibitem[Zhao et~al., 2019]{zhao2019sensitivity}
Zhao, Q., Small, D.~S., and Bhattacharya, B.~B. (2019).
\newblock Sensitivity analysis for inverse probability weighting estimators via
  the percentile bootstrap.
\newblock {\em Journal of the Royal Statistical Society: Series B (Statistical
  Methodology)}, 81(4):735--761.

\bibitem[Zheng et~al., 2021]{zheng2021copula}
Zheng, J., D'Amour, A., and Franks, A. (2021).
\newblock Copula-based sensitivity analysis for multi-treatment causal
  inference with unobserved confounding.
\newblock {\em arXiv preprint arXiv:2102.09412}.

\end{thebibliography}

\end{document}